\documentstyle{mn}
\input{psfig}

\def\littleprime{\ifmmode{\scriptscriptstyle \prime }
    \else{\hbox{$\scriptscriptstyle \prime$ }}\fi}
\def\littlecirc{\ifmmode{\scriptscriptstyle \circ }
    \else{\hbox{$\scriptscriptstyle \circ $ }}\fi}
\def\littless{\ifmmode{\scriptscriptstyle s }
    \else{\hbox{$\scriptscriptstyle s $ }}\fi}
\def\arcsec{\raise .9ex \hbox{\littleprime\hskip-3pt\littleprime}}
\def\arcmin{\raise .9ex \hbox{\littleprime}}
\def\degree{\raise .9ex \hbox{\littlecirc}}
\def\arcsecpoint{\hbox to 1pt{}\rlap{\arcsec}.\hbox to 2pt{}}
\def\arcminpoint{\hbox to 1pt{}\rlap{\arcmin}.\hbox to 2pt{}}
\def\degreepoint{\hbox to 1pt{}\rlap{\degree}.\hbox to 2pt{}}
\def\gtapr {\lower .1ex\hbox{\rlap{\raise .6ex\hbox{\hskip .3ex
        {\ifmmode{\scriptscriptstyle >}\else
                {$\scriptscriptstyle >$}\fi}}}
        \kern -.4ex{\ifmmode{\scriptscriptstyle \sim}\else
                {$\scriptscriptstyle\sim$}\fi}}}
\def\ltapr {\lower .1ex\hbox{\rlap{\raise .6ex\hbox{\hskip .3ex
        {\ifmmode{\scriptscriptstyle <}\else    
                {$\scriptscriptstyle <$}\fi}}}
        \kern -.4ex{\ifmmode{\scriptscriptstyle \sim}\else
                {$\scriptscriptstyle\sim$}\fi}}}
\title{MERLIN observations of Stephan's Quintet}
\author[E. Xanthopoulos et al.] {E.~Xanthopoulos$^{1,2,3}$,
T.~W.~B.~Muxlow$^{3}$, P.~Thomasson$^{3}$, S.~T.~Garrington$^{3}$ \\
$^{1}$University of California Davis, Department of Physics, Davis, CA 95616  \\
$^{2}$IGPP/Lawrence Livermore National Laboratory, Livermore, CA 94550 \\
$^{3}$University of Manchester, Jodrell Bank Observatory, Macclesfield, Cheshire SK11 9DL, England}

\date{Accepted  .
     Received   }

\begin{document}
\maketitle

\begin{abstract}

We present MERLIN L-band images of the compact galaxy group,
Stephan's Quintet. The Seyfert 2 galaxy, NGC 7319, the brightest
member of the compact group, is seen to have a triple radio
structure typical of many extra-galactic radio sources which have
a flat spectrum core and two steep spectrum lobes with hot spots.
The two lobes are asymmetrically distributed on opposite sides of
the core along the minor axis of the galaxy. Ultraviolet emission
revealed in a high resolution HRC/ACS HST image is strongly
aligned with the radio plasma and we interpret the intense star
formation in the core and north lobe as an event induced by the
collision of the north radio jet with over-dense ambient material.
In addition, a re-mapping of archive VLA L-band observations
reveals more extended emission along the major axis of the galaxy
which is aligned with the optical axis. Images formed from the
combined MERLIN and archive VLA data reveal more detailed
structure of the two lobes and hot spots.


\end{abstract}


\footnotetext{Contact e-mail: exanthop@igpp.ucllnl.org}

\section{Introduction}
\begin{figure}
\begin{center}
\setlength{\unitlength}{1cm}
\begin{picture}(8,8)
\put(0.3,0.2){\includegraphics{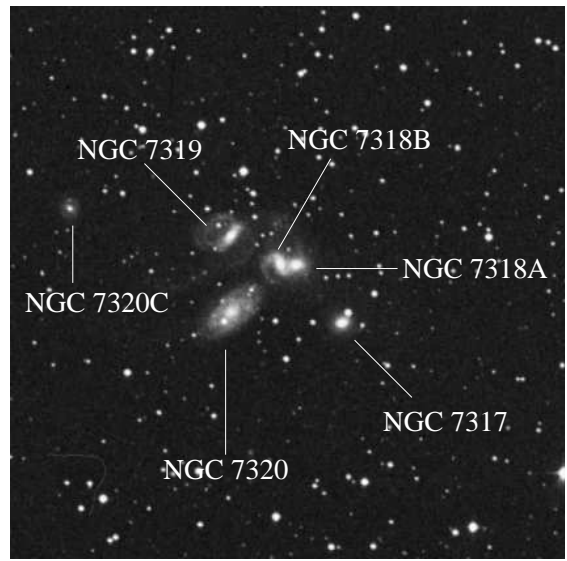}}
\end{picture}
\caption{Digital Sky Survey map covering the member galaxies of
Stephan's Quintet.} \label{names}
\end{center}
\end{figure}
Stephan's Quintet (also known as Arp319 and VV228), which can be
regarded as the prototype of the compact group class of galaxies
(H92 in the Hickson 1982 catalog), was the first to be discovered
120 years ago by Stephan (1877). It has been studied intensively
in almost every wavelength domain and it is a remarkable example
of a group of galaxies visited by infalling neighbours.

There has been much controversy regarding the redshifts of its galaxies
and their nature. The members of the group (Fig.~\ref{names})
were initially thought to be NGC 7317, NGC 7318A, NGC 7318B,
NGC 7319, NGC 7320 and NGC 7320C.
However, it was later found that NGC 7318B had a velocity almost
1000 km s$^{-1}$ lower than the mean redshift of the group ($\approx$6600 km s$^{-1}$)
and NGC 7320 was at a redshift of only $\approx$800 km s$^{-1}$.
Moles et al. (1997) have argued that the discordant-redshift galaxy,
NGC 7320, is an unrelated foreground galaxy and have suggested that the kernel of three galaxies, NGC 7317,
NGC 7318A and NGC 7319, has been visited by
two ``intruders" (NGC 7320C and NGC 7318B) at different times in the past few 10$^{8}$ years.
As a result of these ``intrusions", the
intragroup medium has been filled with HI gas, stripped from the spiral members of the group
(Allen \& Sullivan 1980; Sulentic \& Arp 1983; Shostak et al. 1984; Pietsch et al. 1997; Williams 1998).

\begin{figure}
\setlength{\unitlength}{1cm}
\begin{picture}(9,7)
\put(0.2,0){\includegraphics{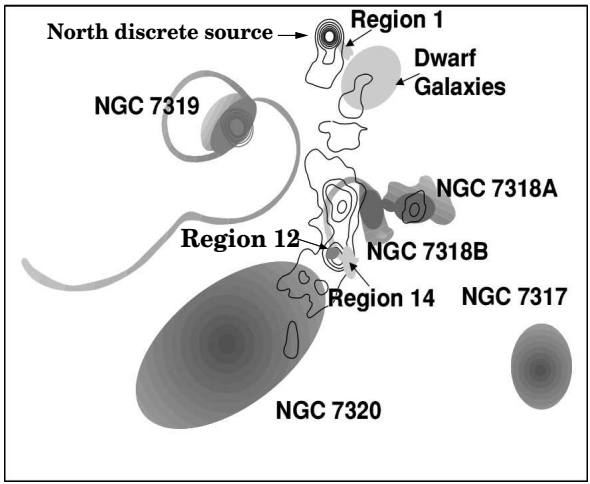}}
\end{picture}
\caption{Schematic showing the relative positions of the radio
emission detected by van der Hulst \& Rots (1981) and the
optically visible galaxies. Also shown are the locations of a
group of dwarf galaxies and regions of H$\alpha$ emission,
designated Regions 1, 12 and 14 by Mendes de Oliveira et al.
(2001). Both the dwarf galaxies and H$\alpha$ emission are
considered to be associated with NGC 7318B.} \label{schema}

\end{figure}
Allen \& Hartsuiker (1972) produced the first continuum radio
image of Stephan's Quintet at a wavelength of 21 cm using the
Westerbork Synthesis Radio Telescope. At a resolution of $\approx$
20\arcsec $\times$ 45\arcsec, it showed an unresolved source
coincident with NGC 7319, and an unusual semi-circular arc of
emission which appeared to envelop it. Since then, VLA
observations at 20 cm wavelength by van der Hulst \& Rots (1981)
with a resolution of 6\arcsec have shown more clearly the main
components of the radio emission associated with Stephan's
Quintet. These are shown in the schematic in Fig.~\ref{schema}.
More recent VLA observations at 3.6 cm, 6 cm and 20 cm by Aoki et
al. (1999) show that the source centered on NGC 7319 has three
compact components, aligned perpendicular to the disc of the
galaxy and embedded in more diffuse emission. According to Aoki et
al. (1999), all the components of the emission have steep spectra
which, they indicate, is commonly found in Seyfert galaxies. Van
der Hulst \& Rots (1981) resolved the arc of emission between NGC
7319 and NGC 7318B into distinct sources embedded in more diffuse
emission which is partly coincident with spiral-arm-like features
in NGC 7318B. They considered this arc of emission to be more like
a letter ``W" on its side, associating it with the starburst
region. They also noted a discrete, unresolved source, just to the
north of the ``W", which, with no optical counterpart, they
considered to be an unrelated background object.

In this paper, high resolution MERLIN radio observations at L band
(18 cm) are presented for the first time. Also presented are the
results of a re-analysis of L band (20 cm) VLA archive data and a
combination of this and the MERLIN data to reveal in much more
detail the structure of the radio emission, not only of NGC 7319,
but of the hitherto unresolved sources in the field. An archival
near-UV high resolution HST image of NGC 7319 is also analyzed and
the UV, optical and radio emission correlation is examined and
discussed. In Section 2, the MERLIN and HST observations and data
analysis as well as a re-analysis of the VLA data are described.
Maps and results are included in Section 3 and the paper finishes
with a discussion and conclusions. Throughout this paper we use
H$_{0}=$ 65 km s$^{-1}$ Mpc$^{-1}$ unless otherwise specified. 1\arcsec
corresponds to 485 pc at the redshift z=0.02251 of NGC 7319.

\section{Observations \& Data Reduction}
\subsection{MERLIN data}

\begin{figure*}
\begin{center}
\setlength{\unitlength}{1cm}
\begin{picture}(8,14)
\put(-4,-1.5){\includegraphics{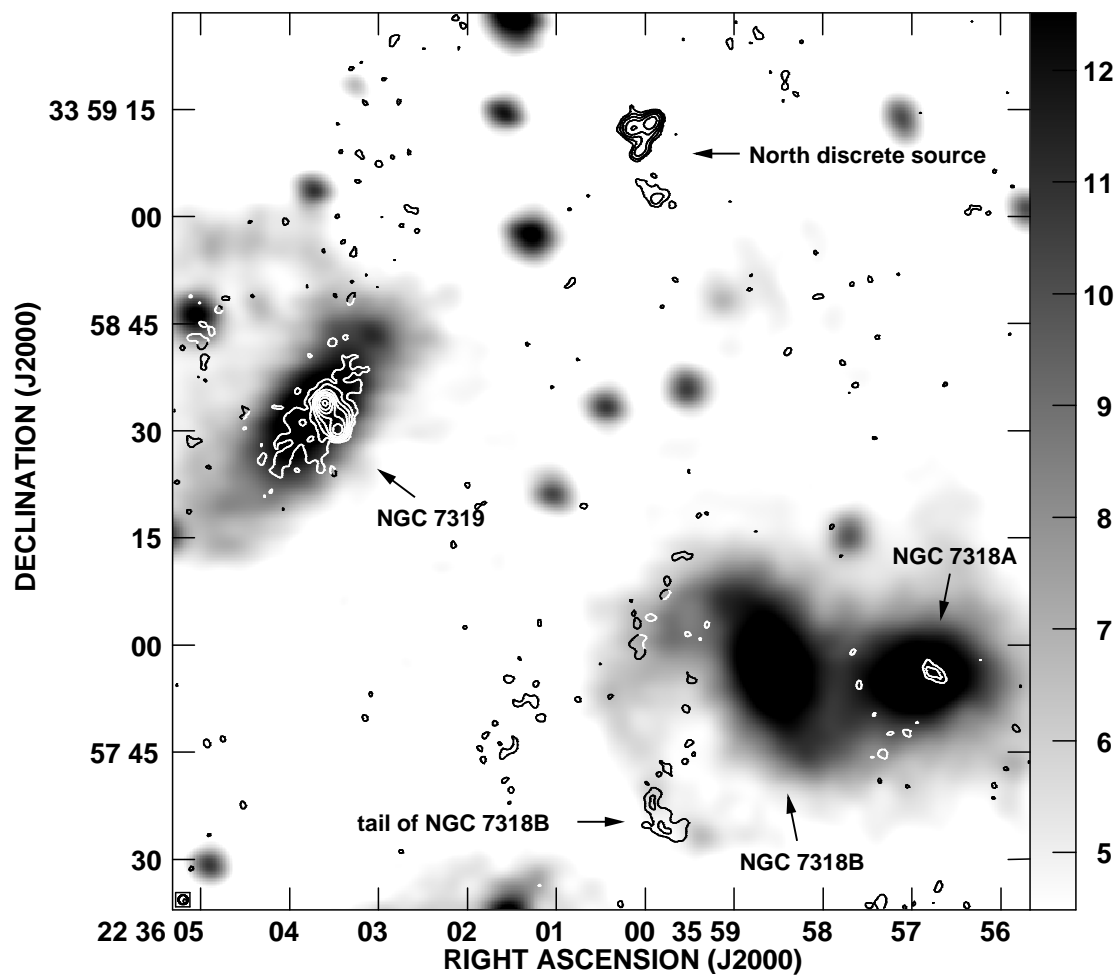}}
\end{picture}
\caption{20 cm naturally weighted VLA archive data image of
Stephan's Quintet at a resolution of 1\arcsecpoint3 overlaid on a
Digital Sky Survey image. Contour levels are at
6.97e-5$\times$(1,2,4,8,16,32,64,128,256) Jy/beam. The peak flux
is 10.3 mJy/beam and the rms noise level is $\sim$27
$\mu$Jy/beam.}\label{vla_tot}
\end{center}
\end{figure*}

Observations of Stephan's Quintet were made with MERLIN at L band
in April and June 1999. A bandwidth of 16 MHz in both left and
right circular polarizations was used, centered on 1658.0 MHz. The
observations were made in ``wide-field" mode in order to cover as
much as possible of the environment of Stephan's Quintet. The
total extent of Stephan's Quintet is $\approx$ 6 $\times$ 6
arcmin$^{2}$ which is easily accessible to MERLIN in its ``wide
field" mode with a field of view at L band of $\approx$ 20
$\times$ 20 arcmin$^{2}$, though with image-smearing towards the
edges of the field. In this mode, the cross-hands of polarization
(LR and RL) are not correlated in order to increase by a factor of
2 the number of channels into which the 16 MHz LL and RR
correlations can be divided; i.e. 32 channels of 0.5 MHz
bandwidth. By preserving these 0.5 MHz bandwidth channels in the
subsequent processing, it is this 0.5 MHz and not the full 16 MHz
bandwidth correlated that determines the field of view. Thus the
angular radius from the pointing centre at which a 10\% image
smearing due to the 0.5 MHz `bandwidth' occurs is $\approx$ 4.7
arcminutes. The integration time per data point was 4 seconds,
which would give rise to a 10\% integration time smearing at an
angular distance from the pointing centre of $\approx$ 5.3
arcminutes. The pointing centre for the MERLIN observations was at
$\alpha =$22$^{h}$36$^{m}$00.51$^{s}$ and $\delta
=$33$^\circ$58$\arcmin$03$\arcsecpoint$76 (J2000). Observations of
Stephan's Quintet were interleaved with those of a nearby phase
calibration source, 2243+357, with a total cycle time (including
telescope drive times) of 7.5 min (5.5 min on source, 2 min on the
calibrator). Although the total on source time was $\approx$ 70
hours, one or more telescopes were lost for significant periods of
time during the observations because of weather conditions; i.e.
wind.

The data were edited, corrected
for elevation-dependent effects, non-closing baseline errors and bandpass response
using the standard MERLIN analysis programs. The flux calibration was established from observations of
3C286, whose flux was computed to be 13.639 Jy at 1658 MHz on both the scale of Baars et al. (1977)
and the VLA calibration list. The telescope relative sensitivities, bandpass and non-closing
corrections were obtained from observations of the point source OQ208, whose flux density was
determined to be 1.18 Jy from a comparison of its signal amplitude with that from 3C286 on the shortest
MERLIN baseline.

After combining the data from the different days, further processing was carried out
using the MERLIN automated pipeline procedure, which uses the NRAO {\sc aips} tasks.
In this, the target source (Stephan's Quintet) was mapped after its data had been corrected
for further amplitude and phase errors derived as a result of mapping the phase-calibration
source, 2243+357. Finally, those fields within the MERLIN primary beam showing radio emission
were carefully mapped using different data weighting functions to produce the final images.

\subsection{VLA data}
Although the observations of Aoki et al. (1999) were directed
solely at the Seyfert galaxy, NGC 7319, within Stephan's Quintet,
it has been possible to use the data from their 20 cm observations
on 4 November 1996 (VLA programme AA206) to produce a 20 cm image
of a much wider field. The data were obtained from the VLA
archive, together with corresponding observations of the flux and
baseline calibrator, 3C48, and a phase reference source, 2236+286.
The flux density of 3C48 in the two IFs centered at 1364.9 MHz and
1435.1 MHz were taken to be 15.617 Jy and 16.319 Jy respectively,
based on the VLA calibration scale. As for the MERLIN
observations, after initial calibration using 3C48, the data were
corrected for additional amplitude and phase errors derived as a
result of mapping the phase calibration source. The images
obtained solely from the VLA data, were produced using the NRAO
{\sc aips} tasks.

Finally, after ensuring correct positional and flux scale alignment, images of
those fields showing radio emission within the primary beam of the telescope were
produced from the combined MERLIN+VLA data. The maps for the northern
discrete source and for NGC 7319 are shown in Fig.~\ref{vlamerlin} and Fig.~\ref{NGC7319}
respectively. The resolution of the combined data is 0\arcsecpoint15 and the rms noise level is
30 $\mu$Jy/beam. Although the core emission (component B) of NGC 7319 is now less obvious
than in the MERLIN only image as it blends into the extended emission, the overall structure
of NGC 7319 and, in particular, the structure of the previously unidentified northern source
are now much more clear.

\subsection{HST/ACS data}

An HST image of NGC 7319 was obtained in 2002 with the High
Resolution Channel (HRC) of the ACS camera as part of the HST
proposal 9379 on Near Ultraviolet imaging of Seyfert galaxies. The
image was taken through the F330W (HRC U) filter (effective
wavelength 3354\AA /\ 588\AA) with a 1200 sec exposure. The HRC
has a 29\arcsec $\times$ 25\arcsec field of view and a plate scale
of $\sim$0.027 \arcsec\/pixel. Basic two-dimensional image
reductions (overscan, bias, dark subtraction and flat fielding) as
well as cosmic ray rejections from the CR-SPLIT=2 data were
performed with the CALACS pipeline processing. The observations
used the dither patterns and so further processing by PyDrizzle
corrected for geometric distortion of the ACS camera and combined
the dithered images into one final reduced and calibrated image.
Using IRAF we removed any remaining cosmic ray events from the
data and rotated the image to the cardinal orientation (North up
and East to the left) by means of the keyword ``ORIENTAT" in the
data header. At the redshift z=0.0225 of the galaxy, 1 pixel
corresponds to 13.1 pc for H$_{0}=$ 65 km s$^{-1}$ Mpc$^{-1}$.

The final calibrated reduced ACS image is shown in Fig.~\ref{ACS}.
The Ultraviolet observation through the F330W filter is the optimal
configuration to detect faint star forming regions around the nuclei.
The image, calibrated in counts in e$^{-}$, was multiplied by the
PHOTFLAM keyword from the image header and converted to erg
cm$^{-2}$\AA$^{-1}$. A U magnitude of 16.01 was measured
using the zero point, PHOTZPT.

\section{Maps and Results}

\begin{figure}
\begin{center}
\setlength{\unitlength}{1cm}
\begin{picture}(10,11)
\put(-1.8,-2){\includegraphics{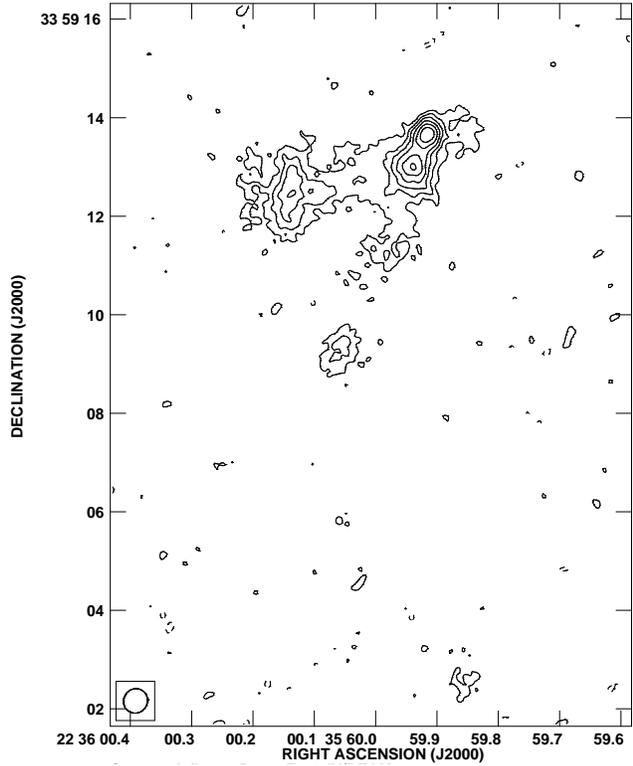}}
\end{picture}
\caption{VLA+MERLIN combined L band map (resolution 0\arcsecpoint15) of the north discrete source.
The contour levels have been set to 7.59e-5(-1,1,2,3,4,5,6,7,8,9,10) Jy/beam. The peak flux is 527
$\mu$Jy/beam and the rms noise level is $\sim$30$\mu$Jy/beam. }
\label{vlamerlin}
\end{center}
\end{figure}

\subsection{\bf Stephan Quintet's radio detections}

The resulting naturally weighted VLA map of
Stephan's Quintet with a beam size of 1\arcsecpoint3 is shown in
Fig.~\ref{vla_tot} (rms noise level 27 $\mu$Jy/beam). The total
intensity contour map is overlaid of the optical Digital Sky Survey
image of the same area. The lowest believable contour is at a level
of approximately a factor of 2 lower than that in the published map of NGC 7319.
The improvement
in this image is probably due in part to taking into account all
of the radio emission in the field in the `reconstruction' of the
image from the data. As might be expected, in comparison with the
MERLIN images, this shows the extended structure more clearly,
though even at this resolution, much of the north-south arc of
emission is considerably resolved and the emission that is visible
in the region at $\alpha \approx$ 22$^{h}$35$^{m}$59.8$^{s}$,
$\delta \approx$ 33$^\circ$57$\arcmin$34$\arcsec$ (J2000) is at
the limit of sensitivity. However, the northern unidentified
source is clearly visible, though somewhat resolved. Also visible
in the image is radio emission at a position corresponding to the
core of NGC 7318A. More specifically the radio detections in the
VLA image are:

\noindent
{\bf a) NGC 7318A:} The two elliptical contours seen at
$\alpha =$ 22$^{h}$35$^{m}$56.8$^{s}$ , $\delta =$ 33$^\circ$57$\arcmin$55$\arcsec$ (J2000)
(Fig.~\ref{vla_tot}), correspond to the centre of NGC 7318A. This is the weak radio detection
also mentioned by van der Hulst\& Rots (1981), Menon (1995), Williams et al. (2002) and
Xu et al. (2003) who measured 1.3, 0.8, 1.4 and 0.95 mJy, respectively.
Our measured flux density is slightly lower, $\approx$0.61$\pm$01 mJy. Bushouse (1987)
reported detection of modest H$\alpha$ emission from the nuclear region of NGC 7318A although more
recently Moles et al. (1998) did not detect any emission from NGC 7318A using longslit spectroscopy.
NGC 7318A is completely resolved by our MERLIN only observations.

\noindent
{\bf b) Tails of NGC 7318B:} The radio emission at the map sensitivity limit seen at the southerly position
$\alpha =$ 22$^{h}$35$^{m}$59.8$^{s}$, $\delta =$ 33$^\circ$57$\arcmin$34$\arcsec$ (J2000) in Fig.~\ref{vla_tot}
is all that remains at this resolution of the arc of emission, probably related to
the southern tail of NGC 7318B and in particular to the bright emission-line regions 12 and 14 as labelled
by Mendes de Oliveira et al. (2001). These regions are examples of non-rotating structures, possibly
associated with HII regions and have star formation rates of $\sim$0.1 M$_{\odot}$ yr$^{-1}$. They are
regions rich in cold and ionized gas.

\noindent {\bf c) The North Discrete Source:} The unidentified
source seen at the northern end of the arc of radio emission
between NGC 7319 and NGC 7318B was first mentioned by van der
Hulst \& Rots (1981), who could not, however, unambiguously say
whether this source was associated with Stephan's Quintet or
whether it was an unrelated background source. The MERLIN+VLA
image of it is shown in (Fig.~\ref{vlamerlin}). The image
indicates that the source is still something of an enigma. Xu et
al. (2003) identify the binary radio source in VLA 1.4 and 4.86
GHz images as the cosmologically distant background source seen in
projection behind SQ (van der Hulst \& Rots 1981, Williams et al.
2002), and note that there is no optical counterpart brighter than
29 mag arcsec$^{-2}$ (Williams et al. 2002), nor any IR
counterpart in the ISO images (Xu et al. 1999, Sulentic et al. 2001).
The total measured flux density
from our radio images, including the southernmost component at
$\alpha =$ 22$^{h}$35$^{m}$59.85$^{s}$, $\delta =$
33$^\circ$59$\arcmin$02$\arcsec$ (J2000), is 7.16$\pm$0.5 mJy. This
flux is less than the 9 mJy flux measured by van der Hulst \& Rots
(1981), and 10.3 mJy measured by Xu et al. (2003) and Williams et
al. (2002), indicating that more extended weak emission has been
resolved.

Although it has been assumed that this is a background source, it
seems a remarkable coincidence that it almost appears to be a
continuation of the extended arc of emission thought to be
associated with NGC 7318B and therefore with Stephan's Quintet. In
the NVSS image of the region with an angular resolution of
45\arcsec, it can be seen that there is extended emission to the
northwest of NGC 7319 which extends beyond and to the south-west
of the unidentified source position. The peak of this extended
continuum emission does not coincide with the unidentified source,
but does encompass the dwarf galaxies and emission line regions
(1A and 1B) associated with NGC 7318B as found by Mendes de
Oliveira et al. (2001) (indicated as Dwarf Galaxies and Region 1
in Fig.~\ref{schema}). The peak of the radio emission seen in the
MERLIN+VLA image lies just to the north-east of Region 1. The
positional offset is similar to that of the radio emission (still
visible in our new VLA image at the southern end of the arc) from
the bright emission line region 14 found by Mendes de Oliveira et
al. (2001).

\subsection{\bf NGC7319}
\subsubsection{Radio}
\begin{figure}
\begin{center}
\setlength{\unitlength}{1cm}
\begin{picture}(5,10)
\put(-1.8,-2){\includegraphics{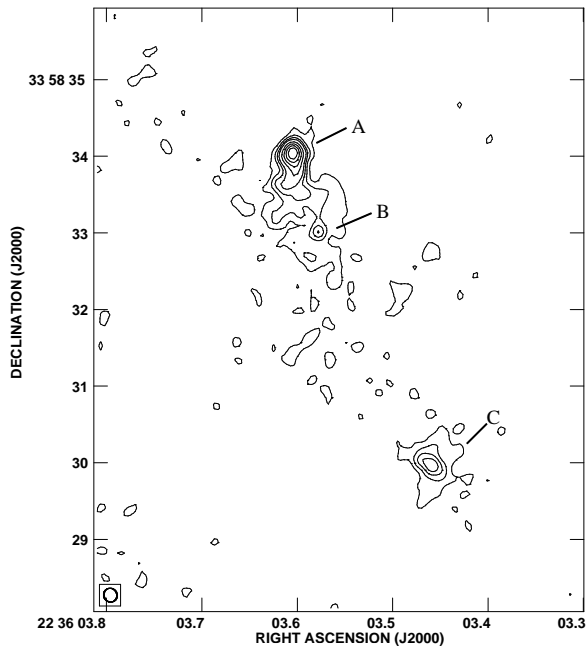}}
\end{picture}
\caption{20 cm naturally weighted MERLIN image of NGC 7319
(resolution 0\arcsecpoint15). The compact components are labelled
A, B and C. Contour levels are set to
1.22e-4$\times$(-1,1,2,4,8,16,32) Jy/beam. The peak flux is 1.84
mJy/beam and the rms noise level is $\sim$47 $\mu$Jy/beam.}
\label{NGC7319m}
\end{center}
\end{figure}

\begin{figure}
\begin{center}
\setlength{\unitlength}{1cm}
\begin{picture}(8,8)
\put(-1.8,-2){\includegraphics{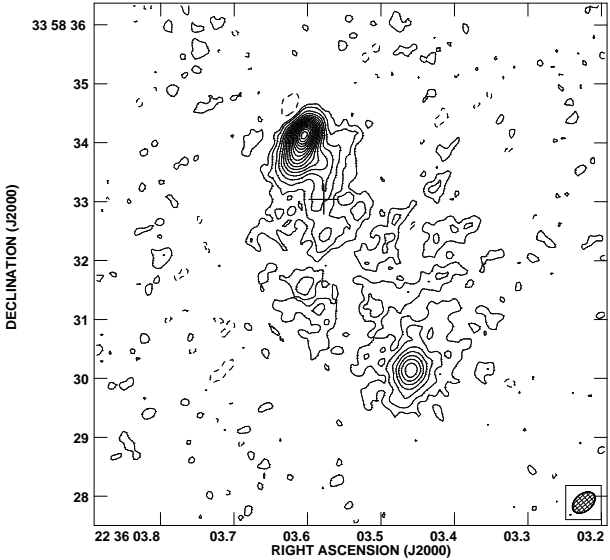}}
\end{picture}
\caption{VLA+MERLIN combined L band map of NGC 7319 (resolution 0\arcsecpoint15).
The contour levels have been set to
7.59e-5$\times$(-1,1,2,4,6,8,10,12,14,16,18,20,22,24,26,28,30,32,34,36,38,40). The peak flux is 3.25 mJy/beam and the rms noise level is $\sim$29
$\mu$Jy/beam. The cross indicates the position of the B component (core) as labelled in
Fig.~\ref{NGC7319m}.}
\label{NGC7319}
\end{center}
\end{figure}

Aoki et al. (1999) have indicated that the previously known
compact components (labelled A, B and C in Fig.~\ref{NGC7319m} after Aoki et al. 1999)
and diffuse emission of NGC 7319 all have steep spectra. However,
our MERLIN only 20 cm image (Fig.~\ref{NGC7319m}), which resolves
out much of the extended emission, indicates component B to be
unresolved and to have a flux 1.07 mJy, which is very comparable
with the value of 1 mJy quoted by Aoki et al. (1999) for its 6 cm
flux. At the same time the flux densities of components A and C
shown in Table~\ref{fluxtab}, although lower than the 20 cm VLA
fluxes measured by Aoki et al. (1999), are more than double in
flux than the 6 cm flux densities quoted by Aoki et al. (1999)
(see their Table~1). Component B then would appear to have a flat
spectrum, and as such is considered to be the core of the galaxy.
We are also able to measure a size for component B. This could not
be defined by Aoki et al. (1999) due to the closeness of
components A and B, and indeed a merging of these components in
their 20 cm map. The positions, flux densities, and sizes (FWHM)
of the three compact components were measured using the {\sc aips} task
{\sc jmfit}, a two-dimensional elliptical Gaussian fitting program.
These parameters are given in Table~\ref{fluxtab}. The sizes are
given after deconvolution from the beams.

The radio structure of NGC7319 looks like a small, asymmetric
FRII radio source, with a flat-spectrum core and two extended lobes,
both containing compact hot-spots.  There is no evidence of a jet
as is often the case. However, the size, luminosity and
spiral host galaxy of NGC 7319 is consistent with it being a Seyfert 2 galaxy (Durret 1987).
The asymmetric arm ratio of 1:3.1 for the distances between
the core (component B) and the two hotspots (components A and C respectively) possibly indicates
that material to the north-east is blocking the outflow of plasma in that direction whereas plasma
outflow to the south-west is much less inhibited.

High resolution radio surveys of Seyfert galaxies (eg. Thean et al.
2000; Thean et al. 2001 and references therein) show
that Seyfert 1 and 2 galaxies have a range of radio structures and sizes.
At 0\arcsecpoint25 resolution, the majority are unresolved or only
marginally resolved, but approximately 10\% show double structures
and a similar fraction show triple or multi-component linear
structures.

\begin{figure}
\begin{center}
\psfig{file=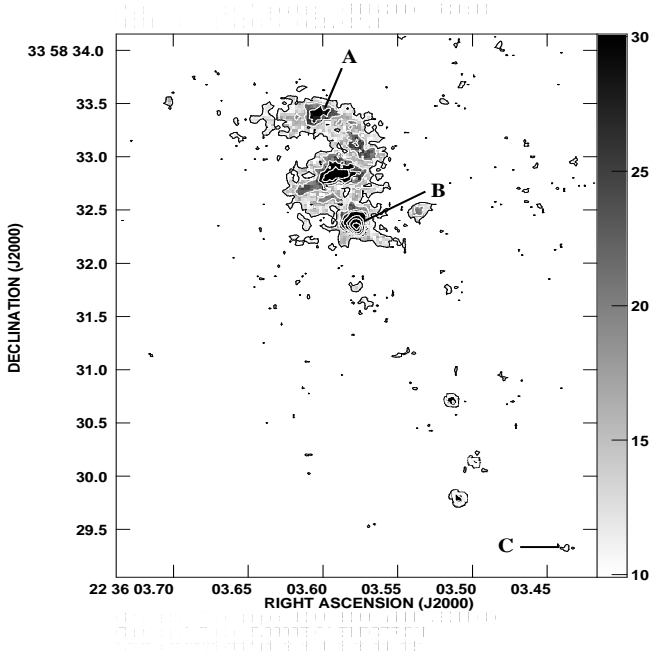,width=3.5in,height=3.5in}
\caption{HST ACS/HRC UV (F330W) image of NGC 7319 (resolution
0\arcsecpoint027). Components A, B and C are indicated in the image.
North is up and East to the left. The sidebar shows the relative
intensity of the greyscale image. Contour levels are overlaid to accentuate features and
are set at 2.0e-20$\times$(1,2,4,6,8,10,12) erg cm$^{-2}$\AA$^{-1}$.}
\label{ACS}
\end{center}
\end{figure}

NGC 7319 has a radio luminosity typical of a Seyfert 2 galaxy and a linear size
at the upper end of the size distribution. It's FRII-like
(such as Cygnus A - which have very well collimated jets ending at high surface brightness hotspots
and large diffuse lobes) structure is similar to NGC 5929 (Su et al. 1996),
Mrk 463 (Kukula et al. 1999) or IC 5063 (Morganti et al. 1998).
The identification of a growing number of Seyferts
with a simple FRII-like triple structure is providing increasing evidence
for the presence of shocks due to the interaction of radio jets with
the ISM in the inner few kpc and calling into question the usual assumption
that the compact radio components of Seyfert galaxies, which often
closely correspond to peaks in the NLR emission, represent the galaxy nucleus.
If NGC 7319 were at twice the distance, components
B and C would probably not be detected and it would be assumed that A is a barely resolved single component
Seyfert galaxy.

In addition to the known three compact components and diffuse
emission aligned along the minor axis of NGC 7319, the new image
formed from the VLA archive data also shows, for the first time,
emission of extent $\approx$20$\arcsec$ perfectly aligned with the
strong bar at a PA = 142$\degree$ (Durret 1987), which is also the
position angle of the major axis of the `optical' Seyfert galaxy.
Assuming a heliocentric systemic velocity of 6740 km s$^{-1}$,
this corresponds to $\sim$ 9.7 kpc for H$_{0}=$ 65 km s$^{-1}$
Mpc$^{-1}$. Being close to the limit of detection, only an
approximate estimate for its flux can be made and this value is
5.6$\pm$0.3 mJy. This weak extended emission, may be associated
with backflow from the lobes of the ``double" radio source.
However, it may also be associated with star formation in the
nuclear region of this interacting galaxy and be unrelated to the
radio lobes. Assuming the latter to be true, and using the
formulae given in Mobasher et al. (1999), SFR$_{1.4}$ =
$\frac{L_{1.4}}{8.07 \times 10^{20} W Hz^{-1}}$ M$_{\odot}$
yr$^{-1}$, it is possible to calculate a star formation rate for
the Seyfert galaxy of 8.4 M$_{\odot}$yr$^{-1}$, from the
monochromatic luminosity at 1.4 GHz $L_{1.4}$ = 6.89 $\times$
$10^{21}W Hz^{-1}$ (once again assuming a value for H$_{0}$ of 65
km s$^{-1}$ Mpc$^{-1}$ and z=0.02251 for NGC 7319). A 10.2$\pm$0.5
mJy flux density was also measured from the VLA data for the
extended emission along the minor axis.

The total MERLIN flux of NGC 7319 is 12.61$\pm$0.5 mJy. This value is much lower than
$\sim$ 28.5$\pm$0.5 mJy measured by Van der Hulst \& Rots (1981), Aoki et al. (1999),
Williams et al. (2002) and Xu et al. (2003). This is expected as in the high resolution MERLIN image,
most of the extended diffuse emission has been completely washed out.
This diffuse emission however is pronounced in
the VLA+MERLIN map (Fig.~\ref{NGC7319}) extending between components A and C. The core,
component B, is now lost in the diffuse emission and we mark its position, as defined from
the MERLIN only map, with a cross in Fig.~\ref{NGC7319}.

\begin{table*}
\caption{MERLIN 20~cm parameters of the compact components in NGC 7319}
\begin{tabular}{lccccc}\hline
          & \multicolumn{2}{c}{Position (J2000)} &  Flux density & Size \\
Component & $\alpha$ & $\delta$ & (mJy) & (arsec) \\ \hline
& & &  \\
Compact A & 22$^{h}$36$^{m}$03.604$^{s}$ & 33$^\circ$58$\arcmin$33$\arcsecpoint$92 & 7.86 $\pm$ 0.35 (mJy) &
0.64 $\times$ 0.23 & \\
Compact B & 22$^{h}$36$^{m}$03.578$^{s}$ & 33$^\circ$58$\arcmin$33$\arcsecpoint$02 & 1.07 $\pm$ 0.16 (mJy) &
 0.23 $\times$ 0.16&\\
Compact C & 22$^{h}$36$^{m}$03.458$^{s}$ & 33$^\circ$58$\arcmin$30$\arcsecpoint$00 & 3.68 $\pm$ 0.42 (mJy)& 0.51
$\times$ 0.39& \\
\end{tabular}
\label{fluxtab}
\end{table*}

\subsubsection{Comparing radio and ultraviolet images}

\begin{figure}
\begin{center}
\setlength{\unitlength}{1cm}
\begin{picture}(5,10)
\put(-1.8,0){\includegraphics{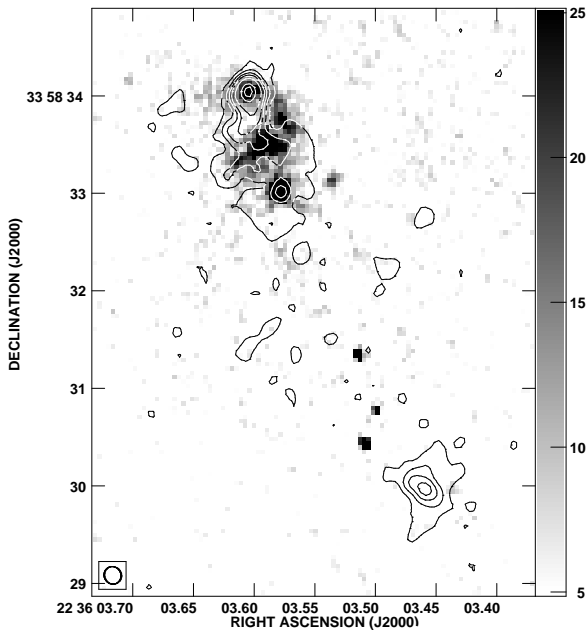}}
\end{picture}
\caption{The 20 cm naturally weighted MERLIN image of NGC 7319
(resolution 0\arcsecpoint15) overlaid on the ACS/HRC UV image. North is up and
East to the left. The sidebar shows the relative intensity of the greyscale image.
The contour levels have been set to 7.59e-5$\times$(1,2,3,4,6,8,10,12).
The peak flux is 1.84 mJy/beam and the rms noise level is $\sim$47 $\mu$Jy/beam. See text for
discussion of aligned features.}
\label{HSTradio}
\end{center}
\end{figure}

It is unfortunate that although the angular resolution of the
MERLIN and HST images is 0\arcsecpoint15 and 0\arcsecpoint027
respectively, due to absolute astrometric uncertainties for HST
data, registration of the two images can not be better than
0\arcsecpoint5. This problem can be mitigated in cases where there is a
bright compact core in the HST image which is reasonably assumed
to be associated with the compact inverted spectrum radio core.
Aoki et al. (1999) attempted to align HST/WFPC2 and VLA data for
NGC 7319 using more accurate astrometric coordinates for the HST
image by smoothing the HST image to ground-based seeing and
assigning an absolute position for the nucleus determined by
ground-based astrometry (Clements 1983). However, the registration
caused a systematic separation of $\sim$0\arcsecpoint7 between
features in HST and radio images which is larger than the
uncertainties in the Clements' position. Due to this and also to
more systematic differences between the optical and radio
astrometric frames they resorted to shift the optical peak onto
the position of the radio component B with a resulting shift in
the HST coordinates of 1\arcsecpoint26.
Since we know that we can not do better than the HST astrometry
we follow the same method of registering the images by shifting
the peak of the UV bright core to the center of the radio nucleus.
This involves a shift of 0\arcsecpoint65 in DEC (the shift in RA is subpixel, 6 mas).
This shift is smaller that the HST astrometric uncertainties. Using
this registration, contours of the MERLIN L-band image are
overlaid on the HST/ACS image in Fig.~\ref{HSTradio}.

\begin{figure}
\begin{center}
\setlength{\unitlength}{1cm}
\begin{picture}(5,10)
\put(-1.8,0){\includegraphics{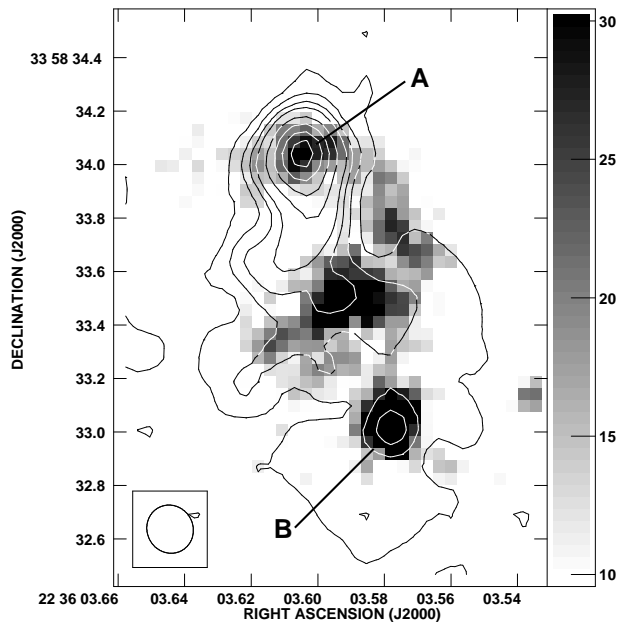}}
\end{picture}
\caption{Enlarged image of Fig.~\ref{HSTradio} isolating the bright core and north lobe.
See text for detailed description.}
\label{HSTradioenl}
\end{center}
\end{figure}

\begin{figure}
\begin{center}
\setlength{\unitlength}{1cm}
\begin{picture}(5,8)
\put(-1.8,0){\includegraphics{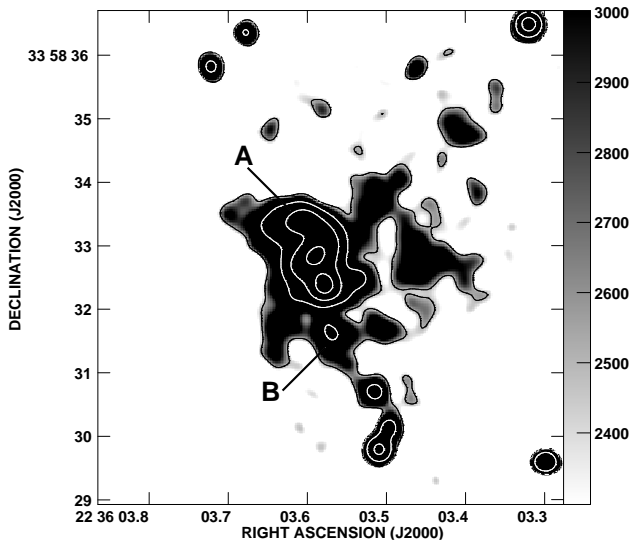}}
\end{picture}
\caption{HST ACS/HRC UV image smoothed to 0\arcsecpoint13 resolution. This enhances and reveals
star formation to the West of NGC 7319.}
\label{HSTsmooth}
\end{center}
\end{figure}

Star formation is evident only in the bright core and the north
lobe. There are only hints of UV emission from component C and
three very compact regions of star formation, north of component C,
trailing the region of the south lobe. The Ultraviolet emission
starts from component B, the core, and has an extent that matches
the north lobe (Fig.~\ref{HSTradioenl}). The emission raps around the north lobe and strong
star formation appears at the position of component A, the end
hotspot of the north lobe. Almost midway between components A and
B, an extended intense star formation region is almost perpendicular
to the radio axis. The UV emission follows closely the elongated contours
(jet-like extension) seen in the MERLIN image below component A.
The SE-NW extent of this UV structure is 0\arcsecpoint85 or
0.45 kpc. A 0\arcsecpoint13 smoothed version of the HST/ACS UV image (Fig.~\ref{HSTsmooth})
brings out an additional star formation region to the W of
this region and just below to what seems now a more marked,
pronounced and extended UV structure between components A and B,
almost arclike in shape. A difference image (the smoothed image
was subtracted from the original unsmoothed image) identifies the
UV emission in the region of the core as the brightest UV region with the
most intense star formation.

\section{Discussion}

Based on the similar properties (strongly interacting galaxies
with tidal tails and bridges, most spirals and S0s, X-ray emitting
intragroup gas with similar X-ray properties) and the HI mass,
Smith et al. (2003) have put forward an evolutionary scenario from
spiral dominated groups to elliptical dominated groups. In this
scheme Stephan's Quintet is probably in an intermediate transition
state between HCG 16 (earlier stage), in which the HI mass is in
both the inner region and tidal features, and NGC 4410 (later
stage), with less HI mass than Stephan's Quintet distributed mostly in
the tidal features. The HI mass of Stephan's Quintet (M$_{HI}$ $\sim$ 10$^{10}$ M$\odot$,
Verdes-Montenegro et al. 2001) is mostly found outside the
galaxies of the group and in tidal features. The galaxies have
almost been completely stripped of their neutral hydrogen (Shostak
et al. 1984, Williams et al. 2002). Interactions between the
galaxies rather than ram pressure have been suggested by
Verdes-Montenegro et al. (2001) to create HI rich tidal features
while depleting the inner regions of galaxies of HI. NGC 4410 is
also the only group of the three that contains a radio galaxy with
radio lobes extending beyond the optical galaxy. The existence of
large radio lobes in the merging
galaxy NGC 4410A shows that such structures can occur fairly early
in the formation of an elliptical, when it is not yet an
elliptical (Smith et al. 2003). Both HCG 16 and Stephan's Quintet
contain Seyfert galaxies. However, in Stephan's Quintet, a
triple radio structure is also present within the dominant galaxy NGC 7319.
Is NGC 7319 an example of a spiral galaxy in which jets
are present but smothered by dense interstellar clouds as
suggested by Wang et al. (2000)? The evidence in this paper seems
to point to this. The jet-cloud interaction is evident in the
arclike-structure seen in NGC 7319 midway between components B and
A. We believe that incidence of the jet on the cloud/dense ambient
gas at this region causes a strong bow-shock that is driven into
the cloud and creates the bright star-formation perpendicular to
the direction of the shock and aligned with the extended radio
contours (see below for UV-radio correlation). The arclike shape
of this structure, as revealed in the smoothed image, is further
support of the presence of bow-shock (e.g. Bicknell et al. 2000).
The jet deposits much of its momentum at this site and it
continues upward to component A where the decelerated jet plasma
accumulates as a radio ``lobe". Wang et al. (2000) present a model
in which weak and intermediate power jets can effectively be
halted or destroyed by reasonably massive clouds and this can be related
to the paucity of extended radio jets in
spiral galaxies. After the jet impacts the cloud, it makes slow
headway and finally is effectively stalled by the cloud. A tilted
interaction area forms in front of the jet nonetheless the
vortical backflow exerts pressure on the jet from the top side,
thus causing the penetrating part of the jet to bent slightly
downward. The curved structure seen at the very end of the UV
continuum can be a result of the bent of the jet as the UV emission trails the
radio emission. Eventually a very extensive cocoon forms on the
top side of the jet and interaction between the cocoon and the jet
induces the jet to evince a substantially wiggly structure.

But is there evidence of such a cloud and dense material in or
near NGC 7319? Observational evidence for dense and cool gas in
high redshift galaxies has been plentiful and includes the
detection of dust, HI, extended line emission, associated
absorption line systems and molecular gas (e.g. De Breuck et al.
2003). NGC 7319 has lost all detectable (upper limit $\sim$10$^{8}
M\odot$) HI, the vast majority of HII regions (expected in a
typical $\sim$SBb spiral) and an uncertain fraction of stellar
mass. At this epoch NGC 7319 is without an ISM that could sustain
the star formation necessary to propagate and define a Population
I spiral pattern. However, a very bright condensation of young
blue stars is found in the northeast edge of the disk and both
this H$\alpha$ and CO emission (Yun et al. 1997) in NGC 7319 is $\sim$8 kpc
north-northeast of the nucleus (6.2$\times$10$^{8} M\odot$). This
region of material is most probably responsible
for confining the North lobe within the spiral structure of NGC 7319. A second
massive CO complex is found within 2 kpc of the nucleus (Yun et
al. 1997), although Gao \& Xu (2000) place this source on the
nucleus. This material coincides with the UV bright core and is
most possibly responsible for the birth of the bow-shock above the
core. The inferred H2 mass for NGC 7319 is 4-7 $\times$ 10$^{9}
M\odot$ (Yun et al. 1997, Gao \& Xu 2000, Smith \& Struck 2001).
The central region of NGC 7319 is luminous from X-ray to radio
wavelengths. It contributes more than half of the MIR/FIR emission
observed from SQ. ISOCAM observations of Stephan's Quintet (Xu et
al. 1999) in the 15 $\mu$m emission, a good nearly extinction-free
star formation rate (SFR) indicator, show the strongest peak
surface brightness on the Seyfert 2 nucleus of NGC 7319 with a
flux of 79.8 mJy.

Alignment of the rest-frame UV continuum emission from the parent
galaxies with the nonthermal radio emission has been discovered in
high-redshift radio galaxies (McCarthy et al. 1987, Chambers et
al. 1987). The nature of this continuum and ``alignment effect''
has remained however unclear. It appears that interactions between
the radio plasma and the ambient gas determine the morphology of
the UV with the radio ejecta sweeping up and compressing the
interstellar medium. In nearby radio galaxies, evidence has been
found for jet-induced star formation, scattered light from hidden
quasar-like AGN and nebular recombination continuum (van Breugel
1985a,b, van Breugel \& Dey 1993, Dey et al. 1996, Tadhunter et
al. 1996, Cimatti et al. 1996). van Breugel et al. (2004) show
that shocks associated with jets may trigger the collapse of
clouds to form stars. Whether this occurs at the impact area, or
along the sides of expanding lobes depends on jet power and the
ambient gas density distribution. It is natural to interpret
enhanced UV continuum as tracing the locus of newly formed stars.
NGC 7319 can most possibly be the first example of jet induced
star formation in a Seyfert galaxy. Up to now there was little
evidence for star formation in the nuclear region of NGC 7319
(e.g. Malkan et al. 1998, Yun et al.1999). A bright UV core is
seen for the first time in Fig.~\ref{ACS}. The size of this
compact emission is 0\arcsecpoint4 or 195 pc. The star formation
appears more pronounced to the N slightly displaced from the
nucleus. Both the orientation and the elongation of this region is
in perfect agreement with the blue elongated region 1\arcsecpoint1
(533 pc) N of the nucleus at a PA of 10\degree, discovered by
Kotilainen (1998) in a B$-$I map of NGC 7319. The UV and blue
emission also agree well with the radio axis as well as with the
[O~III] and the blueward sloping asymmetry detected in the [O~III]
line profiles at the nucleus and northeast of the nucleus (Aoki et
al. 1999). Kotilainen (1998) suggested the blue elongation might
represent scattered light from the nucleus. However, he hinted at
the idea that the blue structures that he saw in Seyfert 2
galaxies might be due to an intrinsically nonstellar continuum,
e.g. emission from high velocity shock waves generated from the
interaction of a radio jet with the extended Narrow Line Region
Gas (e.g. Sutherland 1993). Such an explanation would require a
very close morphological correlation between the continuum and the
high velocity ionized gas which he was not finding. An accurate
knot-to-knot and feature-to-feature matching between the UV
structure and linear radio structure is however established in
Fig.~\ref{HSTradio}. Furthermore we have direct evidence for sharp
boundaries and knotty UV morphology which is the signature of star
forming regions. The extent of the UV emission is the same as the
extent of the radio structure $\sim$5\arcsec (2.42 kpc) and at the
same PA$\sim$25\degree, further supporting close correlation
between jet emission and star formation. In addition, the UV
emission in our Fig.~\ref{ACS} resembles the optical fine
structure, similar to a curved jet, revealed in the difference
WFPC2 (Fig.~5) of Aoki et al (1999)

The absence of any evidence for a polarized, scattered AGN
continuum would support the notion that the active nucleus is not
responsible for the extended UV emission. However, the hard X-ray
emission, peaked on NGC 7319, is consistent with an absorbed
powerlaw, and exhibits a strong Fe K line, providing strong
evidence for an obscured nucleus in NGC 7319 (Awaki et al. 1997).
Optical data (Aoki et al. 1996) also show a strong anisotropic
nuclear ionization radiation which could be produced by an
obscuring torus. On the other hand, the mass of molecular gas in
the central complex observed, is $<$ 2$\times$10$^{8} M\odot$, but
this is inconsistent with the bulk of the observed MIR/FIR
emission originating in a hidden nuclear burst (Xu et al. 1999).
Furthermore, the data show that there is substantial turbulence
produced by a jet-cloud interaction in the vicinity of the radio
components B and A and that the emission lines from the galaxy are
probably related either to the star-forming region or to emission
from radiative cloud shocks rather than excitation by UV-X-ray
emission from the active nucleus. The smoothed image shows further
disturbed UV continuum associated with the extended structure
between components B and A and this disturbance might have
promoted more star formation unveiled in the smoothed image.

In summary, our high resolution radio observations of Stephan's Quintet have shown that
the Seyfert 2 galaxy, NGC 7319, has a flat spectrum core and two very asymmetrically
distributed lobes with hotspots on opposite sides of the core. This adds to the evidence which
calls into question the usual assumption that the compact radio components of Seyfert galaxies
represent the galaxy nucleus. A knot-to-knot morphological correlation between this triple
radio structure and UV emission unveiled in an HST/ACS HRC image of NGC 7319 may suggest
that star formation in this Seyfert galaxy is induced by the collision of a jet with dense
ambient material. We plan to present a detail analysis (star formation rates and kinematical
data) for the UV components in NGC 7319, as well as a proposed  model/simulation
for the jet-cloud interaction in a follow-up paper. We then hope to also
answer the question of whether NGC 7319 will eventually evolve into a classical
double-lobed radio galaxy. To our knowledge, this may well be
the first example of jet induced star-formation in a Seyfert galaxy.

\section*{Acknowledgments}
MERLIN is operated as a National Facility by the Jodrell Bank
Observatory, University of Manchester, on behalf of the UK
Particle Physics \& Astronomy Research Council. The VLA is
operated by the National Radio Astronomy Observatory which is
supported by the National Science Foundation operated under
cooperative agreement by Associated Universities, Inc. This
research was supported by European Commission, TMR Programme,
Research Network Contract ERBFMRXCT96-0034 ``CERES". EXs work was
performed under the auspices of the U.S. Department of Energy,
National Nuclear Security Administration by the University of
California, Lawrence Livermore National Laboratory under contract
No. W-7405-Eng-48 and she also acknowledges support from the National Science
Foundation (grant AST 00-98355). The authors would like to thank the
anonymous referee for helpful suggestions that improved the presentation of
this work.

\end{document}